\documentclass[
 ,final            
  ]
  {aipproc}

\layoutstyle{6x9}

\begin{document}

\title{Higgsless Models: Lessons from Deconstruction\footnote{Presented at the X Mexican Workshop on Particles and Fields, Morelia, Michoacan, Mexico (November 6-12, 2005).} \footnote{MSUHEP-060601}}


\author{E.H. Simmons$^*$, R.S. Chivukula}{
  address={Department of Physics and Astronomy, Michigan State University, East Lansing, MI, USA}
}

\author{H.-J. He}{
  address={Department of Physics, Tsinghua University, Beijing, P.R. China}
 }
 
\author{\\ M. Kurachi}{
  address={Department of Physics and Astronomy, SUNY Stony Brook, NY, USA}
 }
 
\author{M. Tanabashi}{
  address={Department of Physics, Tohoku University, Sendai, Japan}
 }

\begin{abstract}
 This talk reviews recent progress in Higgsless models of electroweak symmetry breaking, and summarizes relevant points of model-building and phenomenology.  
 \end{abstract}

\maketitle


\section{Introduction}

Higgsless models \cite{Csaki:2003dt} do just what their name suggests: they break the electroweak symmetry and unitarize the scattering of longitudinal $W$ and $Z$ bosons without employing a scalar Higgs \cite{Higgs:1964ia} boson.   In a class of well-studied models \cite{Agashe:2003zs,Csaki:2003zu} based on a five-dimensional
$SU(2) \times SU(2) \times U(1)$ gauge theory in a slice of Anti-deSitter space, 
electroweak symmetry breaking is encoded in the boundary conditions of the
gauge fields.  In addition to a massless photon and near-standard $W$ and $Z$ bosons, the spectrum includes an infinite tower of  additional massive vector bosons (the higher
Kaluza-Klein  or $KK$ excitations), whose exchange is responsible for unitarizing longitudinal $W$ and $Z$ boson scattering \cite{SekharChivukula:2001hz,Chivukula:2002ej,Chivukula:2003kq,He:2004zr}. 
The electroweak properties and collider phenomenology of many such models have been discussed in the literature \cite{Cacciapaglia:2004jz,Nomura:2003du,Barbieri:2003pr,Davoudiasl:2003me,Burdman:2003ya,Davoudiasl:2004pw,Barbieri:2004qk,Hewett:2004dv}.

An alternative approach to analyzing the properties of Higgsless models
\cite{Foadi:2003xa,Hirn:2004ze,Casalbuoni:2004id,Chivukula:2004pk,Perelstein:2004sc,Chivukula:2004af,Georgi:2004iy,SekharChivukula:2004mu}
is to use deconstruction 
\cite{Arkani-Hamed:2001ca,Hill:2000mu} and to 
compute the electroweak parameters \cite{Peskin:1992sw,Altarelli:1990zd,Altarelli:1991fk} in a 
related linear moose model \cite{Georgi:1985hf}. We have shown 
\cite{SekharChivukula:2004mu} how to compute 
all four of the leading zero-momentum electroweak parameters defined
by Barbieri et. al. \cite{Barbieri:2004qk} in a very general class of linear moose models.
Using these techniques, we showed \cite{SekharChivukula:2004mu} that a Higgsless model whose fermions are localized (i.e., derive their electroweak properties from a single site on the deconstructed lattice)  cannot simultaneously satisfy unitarity bounds and precision electroweak constraints unless the model includes extra light vector bosons with masses comparable to those of the $W$ or $Z$.

It has recently been proposed \cite{Cacciapaglia:2004rb,Foadi:2004ps,Cacciapaglia:2005pa}
 that the size of corrections to electroweak
processes may be reduced by including delocalized fermions. In deconstruction, a delocalized fermion is realized as a fermion whose $SU(2)$ properties arise from several
sites on the deconstructed lattice \cite{Chivukula:2005bn,Casalbuoni:2005rs}. 
We have studied  \cite{SekharChivukula:2005xm} the properties of deconstructed Higgsless models with fermions whose $SU(2)$ properties arise from delocalization over 
many sites of the deconstructed lattice.    In an arbitrary Higgsless model 
we showed that if the probability distribution
of the delocalized fermions 
is appropriately related to the $W$ wavefunction (a condition we call ``ideal'' delocalization) then deviations in precision electroweak parameters are minimized. In particular, three ($\hat S$, $\hat T$, $W$) of the four leading zero-momentum precision electroweak parameters  defined by Barbieri, et. al. \cite{Barbieri:2004qk} vanish at tree-level.

Because Higgsless models with ideally delocalized fermions have vanishing precision electroweak observables, it is necessary to look elsewhere for experimental signatures of these models \cite{Chivukula:2005ji}\cite{Grojean:2006nn}.  We have computed \cite{Chivukula:2005ji} the form of the triple and quartic gauge
boson vertices in these models.  These constraints were shown to provide lower 
bounds of order a few hundred GeV on the masses of the lightest $KK$ resonances above the 
$W$ and $Z$ bosons.  We have also computed \cite{Chivukula:2005ji} the leading Appelquist-Longhitano coefficients in the electroweak chiral Lagrangian \cite{Appelquist:1980vg,Longhitano:1980tm,Longhitano:1980iz,Appelquist:1980ix,Appelquist:1993ka} for these models.  We also studied the collider phenomenology of the $KK$ resonances in 
models with ideal delocalization; because these resonances are 
fermiophobic, traditional direct collider searches are not sensitive to them and measurements of gauge-boson scattering \cite{Birkedal:2004au} will be needed to find them.  

This talk summarizes some of the recent results for Higgsless models, especially those related to model-building and phenomenology.   Each topic can only be touched on briefly here, and the reader is encouraged to visit the cited literature for more extensive discussion and derivations.

\section{A General Higgsless Model}

\begin{figure}
  \includegraphics[height=.15\textheight]{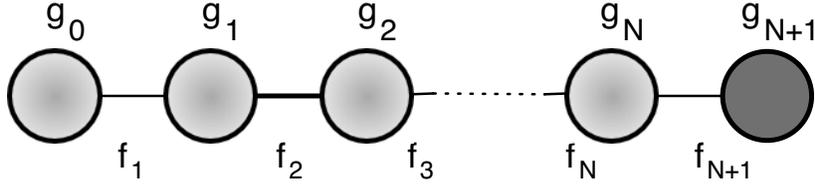}
  \caption{Moose diagram of the type of model analyzed in this talk. Sites $0$ to $N$ are
$SU(2)$ gauge groups, site $N+1$ is a $U(1)$ gauge group.  The gauge couplings and f-constants indicated are left arbitrary in the analysis.  In part of the analysis, the fermions are delocalized
in the sense that the $SU(2)$ couplings of the fermions
arise (potentially) from the gauge groups at all sites from 0 to $N$. The $U(1)$ coupling comes from the
gauge group at site $N+1$.}
\label{fig:tone}
\end{figure}

This discussion focuses on a deconstructed Higgsless model of the general type shown 
 diagrammatically (using ``moose notation'' \cite{Georgi:1985hf}) in fig. \ref{fig:tone}.   The model incorporates an
$SU(2)^{N+1} \times U(1)$ gauge group, and $N+1$ 
nonlinear $(SU(2)\times SU(2))/SU(2)$ sigma models in which the global symmetry groups 
in adjacent sigma models are identified with the corresponding factors of the gauge group.
The Lagrangian for this model at leading order is given by \cite{SekharChivukula:2005xm}
\begin{equation}
  {\cal L}_2 =
  \frac{1}{4} \sum_{j=1}^{N+1} f_j^2 \ \mbox{tr}\left(
    (D_\mu U_j)^\dagger (D^\mu U_j) \right)
  - \sum_{j=0}^{N+1} \frac{1}{2g_j^2} \ \mbox{tr}\left(
    F^j_{\mu\nu} F^{j\mu\nu}
    \right),
\label{lagrangian}
\end{equation}
with
\begin{equation}
  D_\mu U_j = \partial_\mu U_j - i A^{j-1}_\mu U_j 
                               + i U_j A^{j}_\mu,
\end{equation}
where all  gauge fields $A^j_\mu$ $(j=0,1,2,\cdots, N+1)$ are dynamical. The first
$N+1$ gauge fields ($j=0,1,\ldots, N$) correspond to $SU(2)$ gauge groups; the last gauge
field ($j= N+1$) corresponds to the  $U(1)$ gauge group.  The symmetry breaking between
the $A^{N}_\mu$ and $A^{N+1}_\mu$ follows an $SU(2)_L \times SU(2)_R/SU(2)_V$ symmetry
breaking pattern with the $U(1)$ embedded as the $T_3$-generator of $SU(2)_R$.   
Our analysis proceeds for arbitrary values of the gauge couplings and $f$-constants.
In the continuum limit, therefore, this allows for arbitrary background 5-D geometry,
spatially dependent gauge-couplings, and brane kinetic energy terms for the gauge-bosons. 

The neutral vector meson mass-squared matrix is of dimension $(N+2) \times (N+2)$ 
\begin{equation}
M_{Z}^2 = {1\over 4}
\left(
\begin{array}{c|c|c|c|c}
g^2_0 f^2_1& -g_0 g_1 f^2_1 & &  &  \\ \hline
-g_0 g_1 f^2_1  & g^2_1(f^2_1+f^2_2) & -g_1g_2 f^2_2&  &  \\ \hline
 & \ddots & \ddots & \ddots &  \\ \hline
 & & -g^{}_{N-1} g^{}_{N} f^2_{N} & g^2_{N}(f^2_{N} + f^2_{N+1}) & -g^{}_{N} g^{}_{N+1} f^2_{N+1}    \\ \hline
  & & & -g^{}_{N} g^{}_{N+1} f^2_{N+1} & g^2_{N+1}f^2_{N+1} 
\end{array}
\right).
\label{eq:neutralmatrix}
\end{equation}
and the charged current vector bosons' mass-squared matrix is the upper-left $(N+1)  \times (N+1) $ dimensional block of the  $M_{Z}^2$ matrix.
The neutral mass matrix (\ref{eq:neutralmatrix}) 
is of a familiar form that has a vanishing determinant, due to a zero eigenvalue.
Physically, this corresponds to a massless neutral gauge field -- the photon.
The non-zero eigenvalues of $M^2_{Z}$
are labeled by ${\mathsf m}^2_{Zz}$ ($z=0,1,2,\cdots, N$), while
those of ${M}^2_W$ are labeled by ${\mathsf m}^2_{Ww}$ ($w=0, 1,2,\cdots, N$). 

The lowest massive eigenstates corresponding to eigenvalues ${\mathsf m}^2_{Z0}$ and 
${\mathsf m}^2_{W0}$ are, respectively, identified as the usual $Z$ and $W$ bosons.
We will  refer to  these last eigenvalues by their conventional symbols $M^2_Z,\, M^2_{W}$; the distinction between these and the corresponding mass matrices should be clear from
context. 
We will denote the eigenvectors corresponding to the photon, $Z$, and $W$ by
$v^\gamma_i$, $v^Z_i$, and $v^W_j$. These eigenvectors are normalized as 
\begin{equation}
\sum_{i=0}^{N+1} (v^\gamma_i)^2 = \sum_{i=0}^{N+1} (v^Z_i)^2 = \sum_{j=0}^N (v^W_j)^2 = 1~.
\label{eq:evnorm}
\end{equation}
Inspection of the matrix $M^2_Z$ reveals that each component of the photon eigenvector is inversely related to the gauge coupling at the corresponding site
\begin{equation}
v^\gamma_i = {e\over g_i}~,\ \ \ \ {\rm where}\ \ \ \ \ \ {1\over e^2}=\sum_{i=0}^{N+1} {1\over g^2_i}~.
\label{eq:vgamma}
\end{equation}
In the continuum limit, the eigenstates with masses ${\mathsf m}^2_{Ww}$ and 
${\mathsf m}^2_{Zz}$ correspond to the higher Kaluza-Klein (``KK'') excitations of the 
five-dimensional $W$ and $Z$ gauge fields.

\section{\protect{$\hat S$} and Unitarity}

One of the central roles of the Higgs boson in the Standard Model is to unitarize the scattering of electroweak gauge bosons at high energies.  If the Higgs boson were removed from the Standard Model and no new physics were added, $W_L W_L$ spin-0 isospin-0 scattering would violate unitarity at an energy scale of $\sqrt{8\pi}\,$v,  where ${\rm v} = 246$ GeV is the electroweak scale.  In Higgsless models, the Higgs boson is absent, but new physics coupling to the electroweak gauge bosons is present in the form of the higher KK modes of the $W$ and $Z$.  It has been shown that low-energy unitarity of longitudinal electroweak boson scattering is maintained in Higgsless models through exchange of various KK modes  \cite{SekharChivukula:2001hz,Chivukula:2002ej,Chivukula:2003kq,He:2004zr}.  

In \cite{SekharChivukula:2004mu}, we studied a slightly more general deconstructed model than the one shown in Figure 1, a model in which there could be a whole chain of $U(1)$ groups at the right-hand end of the moose, rather than just a single one.  The gauge couplings and $f$-constants were, again, left arbitrary to make our results as broadly applicable as possible. We studied the unitarization of $W_L W_L$ scattering in this model and concluded that KK mode exchange can maintain low-energy unitarity only if the mass-squared $M^2_{W1}$, of the next lightest state after the $W$-boson, is bounded from above by $8\pi {\rm v}^2$.  

We also calculated the form of the corrections to the electroweak interactions for the case in which fermions are localized in the extra dimension; in deconstructed language, the fermions couple to only a single $SU(2)$ group and to a single $U(1)$ group along the moose.  We found \cite{SekharChivukula:2004mu}, that the precision observable $\hat{S}$ (defined below, under Precision Electroweak Corrections) is inversely related to the unitarity-imposed upper bound on $M^2_{W1}$.  In other words, the upper bound on the mass-squared of the first KK mode places a lower bound on $\hat{S}$.  Specifically, we computed 
\begin{equation}
\hat{S} ={1\over 4s^2}\left(\alpha S + 4 c^2 (\Delta \rho - \alpha T) + {\alpha \delta \over c^2}\right) 
\ge {M^2_W\over 8 \pi {\rm v}^2} \simeq 4 \times 10^{-3}~,
\label{eq:final}
\end{equation}
However, this precision observable is constrained by experiment 
\cite{Barbieri:2004qk} to be less than of order $10^{-3}$ in magnitude.  The value predicted by our Higgsless model with localized fermions is significantly disfavored by experiment.

We concluded that Higgsless models with localized fermions  cannot simultaneously satisfy the constraints imposed by precision electroweak corrections and by the need to ensure that scattering of electroweak gauge bosons is unitary.

\section{Ideal Delocalization of Fermions}

Given the problems with localized fermions, it is logical to consider the possibility that
the standard model fermions have wavefunctions with finite extent in the fifth dimension. 
In practice, this means
that the observed fermions are the lightest eigenstates of five-dimensional fermions, just as the $W$ and $Z$ gauge-bosons are the lightest in a tower of ``KK"
excitations. Refs. \cite{Cacciapaglia:2004rb,Foadi:2004ps} show that by adjusting the five-dimensional wavefunction of the
light fermions, one can modify (and potentially eliminate) the dangerously large corrections
to precision electroweak measurements.  Ref. \cite{SekharChivukula:2005xm} took this a step further by introducing ideal fermion delocalization to guarantee that the precision corrections will be small.

 Let us review the basics of fermion delocalization in the language of deconstruction 
 \cite{SekharChivukula:2005xm}. Since a five-dimensional spinor is equivalent to a four-dimensional Dirac fermion, one introduces a separate Dirac fermion at each site ({\it i.e.} one left-handed and one right-handed Weyl spinor per site, $\psi^i_L$ and $\psi^i_R$) on the interior of the moose diagram of fig. \ref{fig:tone}. The chirality of the standard model fermions is introduced by adjusting the boundary conditions for the fermion fields at the ends of the moose. A convenient choice \cite{Cheng:2001nh} 
(consistent with the weak interactions) that we will adopt corresponds to 
\begin{equation}
\psi^{N+1}_L =0~,\qquad \psi^0_R = 0~.
\end{equation}
Discretizing the Dirac action for a five-dimensional fermion in an arbitrary background
metric then corresponds to introducing site-dependent masses ($m_j$) for the Dirac fermions at
each interior site and postition-dependent Yukawa 
interactions ($y_j$) which couple the left-handed modes at site 
$j$ to the right-handed modes at site $j+1$
\begin{equation}
{\cal L}_{5f} = - \sum_{j=1}^{N-1} m_j \bar{\psi}^j_L \psi^j_R 
- \sum_{j=0}^{N-1} f_{j+1}\, y_{j+1}\left( \overline{\psi^j_L} U_{j+1} \psi^{j+1}_R\right) + {h.c.}~,
\label{eq:fmatrix}
\end{equation}
where gauge-invariance dictates that each such interaction include a factor 
of the link field $U_{j+1}$, and we therefore write the corresponding interaction proportional
to $f_{j+1}$.  

Note that in eqn. (\ref{eq:fmatrix}) we have not included a Yukawa coupling corresponding to
link $N+1$.  We will analyze the model in the limit where the lightest fermion eigenstates (which we identify with the standard model fermions) are massless.  
The absence of the  Yukawa couplings at site $N+1$  insures
that the right-handed components of these massless modes are localized 
entirely at site $N+1$. For simplicity, in what follows we will also assume flavor universality, 
{\it i.e.} that the same five-dimensional fermion mass matrix applies to all flavors
of fermions.

In this limit, only the left-handed components of the massless
fermions are delocalized, and their behavior is characterized by a wavefuntion
$|\psi_L \rangle = \left(\alpha_0, 
\alpha_1, \cdots, \alpha_N \right)$ 
where the $\alpha_j$ are complex parameters.  Denoting $\vert \alpha_i\vert^2 \equiv x_i$ and recognizing that  $\sum_{i=0}^N x_i = 1$, we find that the couplings of the ordinary (zero-mode) fermions
in this model may be written
\begin{equation}
{\cal L}_f = \vec{J}^\mu_L \cdot \left( \sum_{i=0}^N x_i \vec{A}^i_\mu \right)
+ J^\mu_Y A^{N+1}_\mu~.
\label{eq:fcoupling}
\end{equation}
As usual, $\vec{J}^\mu_L$ denotes the isotriplet of left-handed 
weak fermion currents and $J^\mu_Y$ is the fermion hypercharge current.

In the presence of fermion delocalization, the coupling of a gauge boson mass eigenstate to a fermion current is the sum of the contributions from each site 
\begin{equation}
g_W = \sum_{j=0}^N x_j g_j v^W_j\ \ \ \ \ \ \ \ \ \ \ \ 
g^W_Z = \sum_{j=0}^N x_j g_j v^Z_j\ \ \ \ \ \ \ \ \ \ \ \ 
g^Y_Z = g_{N+1} v^Z_{N+1}~,
\end{equation}
and therefore reflects the fermion ($x_i$) and gauge boson ($v^W_i,\, v^Z_i$) wave-functions and the site-dependent couplings ($g_i$).

In \cite{SekharChivukula:2005xm}, we  introduced a scheme of 
``ideal fermion delocalization" which guarantees that the precision corrections can be made small. This scheme exploits the fact that the eigenvector for
the lightest massive $W$ and those for each of the $W$ KK modes are mutually orthogonal
\begin{equation}
  \sum_i v_i^{W} v_i^{W_{w}} = 0~.
\end{equation}
Suppose we
choose our ``ideally delocalized" fermion wavefunction $x_i$ to be
related to the form of the $W$ wavefunction 
\begin{equation}
  g_i x_i = g_W  v_i^{W}~,
\label{eq:ideal}
\end{equation}
where the site-independent normalization factor $g_W$ is fixed by the constraint $\sum_i x_i=1$.  Then
the coupling of the fermion to the KK modes vanishes
\begin{equation}
g_{W'} = \sum_{j=0}^N x_j g_j v^{W'}_j =  g^W \sum_{j=0}^N v^W_j v^{W'}_j = 0
\end{equation}
and the KK mode is fermiophobic.    

\section{Precision Electroweak Observables}

To see the implications of ideal delocalization for precision electroweak observables, recall  \cite{Chivukula:2004af} that the most general amplitude for low-energy four-fermion neutral weak current processes in
any ``universal'' model \cite{Barbieri:2004qk} may be written 
as
\begin{eqnarray}
-{\cal M}_{NC} = e^2 {{\cal Q}{\cal Q}' \over Q^2} 
& + &
\frac{(I_3-s^2 {\cal Q}) (I'_3 - s^2 {\cal Q}')}
	{\left({s^2c^2 \over e^2}-{S\over 16\pi}\right)Q^2 +
		{1\over 4 \sqrt{2} G_F}\left(1-\alpha T +{\alpha \delta \over 4 s^2 c^2}\right)
		} 
\label{eq:NC4} \\ \nonumber & \ \ & \\
&+&
\sqrt{2} G_F \,{\alpha \delta\over  s^2 c^2}\, I_3 I'_3 
+ 4 \sqrt{2} G_F  \left( \Delta \rho - \alpha T\right)({\cal Q}-I_3)({\cal Q}'-I_3')~,
\nonumber 
\end{eqnarray}
and the matrix element for charged current process may be written 
\begin{eqnarray}
  - {\cal M}_{\rm CC}
  =  \frac{(I_{+} I'_{-} + I_{-} I'_{+})/2}
             {\left(\frac{s^2}{e^2}-\frac{S}{16\pi}\right)Q^2
             +{1\over 4 \sqrt{2} G_F}\left(1+{\alpha \delta \over 4 s^2 c^2}\right)
            }
        + \sqrt{2} G_F\, {\alpha  \delta\over s^2 c^2} \, {(I_{+} I'_{-} + I_{-} I'_{+}) \over 2}~.
\label{eq:CC3}
\end{eqnarray}
The parameter $s^2$ is defined implicitly in these expressions as the ratio of
the ${\cal Q}$ and $I_3$ couplings of the $Z$ boson.
$\Delta \rho$ corresponds to the deviation from unity of the ratio of the strengths of
low-energy isotriplet weak neutral-current scattering and charged-current scattering.
$S$ and $T$ are the familiar oblique electroweak parameters \cite{Peskin:1992sw,Altarelli:1990zd,Altarelli:1991fk}, 
as determined by examining the {\it on-shell} properties of the $Z$ and $W$ bosons.
The contact interactions proportional to $\alpha \delta$ and ($\Delta \rho - \alpha T$)
correspond to ``universal non-oblique'' corrections arising from the exchange of
heavy KK modes \cite{Chivukula:2004af}.    Hence, if the heavy KK modes are fermiophobic, these electroweak corrections will vanish at leading order.

These "on-shell" paramters may be recast \cite{Chivukula:2004af} as the  the zero-momentum electroweak parameters defined in  \protect{\cite{Barbieri:2004qk} 
\begin{eqnarray}
  \hat{S} &=& \frac{1}{4s^2} \left(
    \alpha S + 4c^2 (\Delta\rho -\alpha T) + \frac{\alpha\delta}{c^2}
  \right), \\
  \hat{T} &=& \Delta\rho, \\
  W &=& \frac{\alpha\delta}{4s^2 c^2}, \\
  Y &=& \frac{c^2}{s^2} (\Delta\rho - \alpha T)~,
\end{eqnarray}
In this language, it can be shown that the electroweak corrections take on a very compact form for ideally delocalized fermions  \cite{SekharChivukula:2005xm}:
\begin{equation}
  \hat{S} = \hat{T} = W = 0, \qquad
  Y = M_W^2 (\Sigma_W - \Sigma_Z)
\end{equation}
where $\Sigma_Z$ and $\Sigma_W$ are the sums over inverse-square masses of the higher neutral- and charged-current KK modes.   While $Y$ is not precisely zero, its value is small (of order $M_W^4 / M_{W'}^4$)

\section{Direct Searches}

Because the electroweak gauge bosons' KK modes are fermiophobic in Higgsless models with ideal fermion delocalization, it will be difficult for direct collider searches to detect the $W'$ and $Z'$ states.
 
Existing searches for $W'$ bosons \cite{PDG} assume that the $W'$ bosons couple to ordinary quarks and leptons (generally with SM strength). All assume that the $W'$ is produced via these couplings; all but one also assume that the $W'$ decays only to fermions and that the $W' \to W Z$ decay channel is  unavailable.   Likewise, existing direct searches for $Z'$ bosons \cite{PDG} rely on $Z' f \bar f$ couplings and assume that the decay channel $Z' \to W^+W^-$ is not available.  None of these searches can constrain the $W_{(n\geq1)}$ and $Z_{(n\geq 1)}$ states of a Higgsless model with ideal fermion delocalization.    Proposed future searches that rely on the $W'$ and $Z'$ couplings to fermions for either production or decay of the KK modes (e.g. \cite{Godfrey:2000hc} , \cite{Godfrey:2000pw} ) will not apply either.  
 
The only way to perform a direct search for the $W_{(n\geq1)}$ and $Z_{(n\geq 1)}$ states will be to study $WW$ or $WZ$ elastic scattering.   If no resonances are seen, these processes will also 
afford the opportunity to constrain the values of the chiral Lagrangian parameters $\alpha_4$ and $\alpha_5$ (see discussion below).

\section{Triple Gauge Vertices}

\begin{figure}
  \includegraphics[height=.25\textheight]{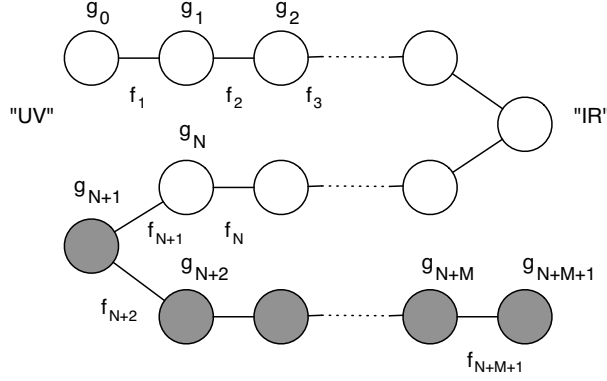}
  \caption{Moose diagram of the deconstructed version of the $SU(2)_A \times SU(2)_B \times U(1)$ 5-dimensional gauge theory in flat space discussed in the sections on Triple Gauge Vertices and Chiral Lagrangian Parameters.  The upper (lower) string of $SU(2)$ groups corresponds to $SU(2)_A$  ($SU(2)_B$)in the continuum.}
\label{fig:two}
\end{figure}

Experimental signatures of Higgsless models with ideally-delocalized fermions will need to rely on couplings among the many electroweak gauge bosons in the theories, since more traditional signatures involving on fermion couplings to the $W'$ and $Z'$ are not available.  One of the best tests of Higgsless models comes from the $WWZ$ vertex.

To leading order, the CP-conserving triple gauge boson vertices
may be written in the Hagiwara-Peccei-Zeppenfeld-Hikasa triple-gauge-vertex 
notation \cite{Hagiwara:1986vm}
\begin{eqnarray}
{\cal L}_{TGV} & = & -ie\frac{c_Z }{s_Z}\left[1+\Delta\kappa_Z\right] W^+_\mu W^-_\nu Z^{\mu\nu}
- ie \left[1+\Delta \kappa_\gamma\right] W^+_\mu W^-_\nu A^{\mu\nu} \cr
&-& i e \frac{c_Z}{s_Z} \left[ 1+\Delta g^Z_1\right](W^{+\mu\nu}W^-_\mu - W^{-\mu\nu} W^+_\mu)Z_\nu 
\label{eq:tgvlag} \\
&-& ie (W^{+\mu\nu}W^-_\mu - W^{-\mu\nu} W^+_\mu) A_\nu~, \nonumber
\end{eqnarray}
where the two-index tensors denote the Lorentz field-strength
tensor of the corresponding field. In the standard model, 
$\Delta\kappa_Z = \Delta\kappa_\gamma = \Delta g^Z_1 \equiv 0$.  The Higgsless models predict $\Delta\kappa_\gamma = 0$ but generally have non-zero values of $\Delta g^Z_1$ and $\Delta\kappa_Z$.

In models with ideally-delocalized fermions, experimental constraints from LEP II measurements of the triple-gauge-boson vertices can provide valuable bounds on the $KK$ masses \cite{Chivukula:2005ji}. 
   The 95\% c.l. upper limit (recalling that $\Delta g^Z_1$ is positive in our models) is $\Delta g^Z_1 \leq 0.028$ \cite{LEPEWWG}.
We can estimate the degree to which this constrains Higgsless models with ideal delocalization by considering how $\Delta g^Z_1$ is related to the mass of the lightest $KK$ resonance.  

For a deconstructed version of an $SU(2)_A \times SU(2)_B \times U(1)$ gauge theory in 5-dimensional flat space (see Figure 2), the form of $\Delta g_1^Z$ is found in 
\cite{Chivukula:2005ji} to be
\begin{equation}
\Delta g^Z_1 =  \Delta \kappa_Z = \frac{\pi^2}{12 c^2} \left(\frac{M_{W}}{M_{W_1}}\right)^2 \left[\frac14 \cdot \frac{7 + \kappa}{1+\kappa}\right]
\end{equation}
where  $\kappa = g^2_{5WB} / g^2_{5WA}$.   Inserting numerical values for $M_W$, and $c$ and denoting the 95\% c.l. experimental 
upper bound on $\Delta g^Z_1$ as $\Delta g_{max}$, we find the bound
\begin{equation}
M_{W_1} \geq 500 {\rm GeV} \sqrt{\frac{0.028}{\Delta g_{max}} \ \left[\frac14 \cdot \frac{7 + \kappa}{1+\kappa}\right]}
\end{equation}
The LEP II data therefore implies a 95\% c.l. lower bound of 500 GeV on the first $KK$ resonance in flat  space models for $\kappa = 1$ and lower bounds of 250 - 650 GeV as $\kappa$ varies from $\infty$ to 0.

\section{Chiral Lagrangian Parameters}
The language of the
effective electroweak chiral Lagrangian is useful for phenomenological studies.  As discussed in \cite{Chivukula:2005ji}, the operators \cite{Appelquist:1980vg,Longhitano:1980tm,Longhitano:1980iz,Appelquist:1980ix,Appelquist:1993ka}  present in our Higgsless models are:
\begin{eqnarray}
{\cal L}_1 &\equiv& \frac12 \alpha_1 g_W g_Y B_{\mu\nu} Tr(T W^{\mu\nu})\\
{\cal L}_2 &\equiv& \frac12 i \alpha_2 g_Y B_{\mu\nu} Tr(T [V^\mu, V^\nu])\\
{\cal L}_3 &\equiv& i \alpha_3 g_W Tr(W_{\mu\nu} [V^\mu, V^\nu])\\
{\cal L}_4 &\equiv& \alpha_4 [ Tr(V^\mu V^\nu)]^2\\
{\cal L}_5 &\equiv& \alpha_5 [Tr(V_\mu V^\mu)]^2~.
\end{eqnarray}
Here $W_{\mu\nu}$, $B_{\mu\nu}$, $T \equiv U\tau_3U^\dagger$ and $V_\mu \equiv (D_\mu U)U^\dagger$, with
$U$ being the nonlinear sigma-model field  arising from $SU(2)_L \otimes SU(2)_R \to
SU(2)_V$, are the $SU(2)_W$-covariant and $U(1)_Y$-invariant building blocks of the expansion.
The relationships of the $\alpha_i$  to several alternative parametrizations are given in \cite{Chivukula:2005ji}; we note here that $S = - 16\pi \alpha_1$ and that the leading corrections to $WW$ and $WZ$ elastic scattering arise from $\alpha_{4,5}$.  The following table shows the values of the Appelquist-Longitano parameters in the deconstructed version of a 5-dimensional $SU(2)_A \times SU(2)_B \times U(1)$ gauge theory in flat space

\begin{table}[ht]
 \begin{tabular}{|c|c|c|}
 \hline
 flat $SU(2) \times SU(2)$ & & \\
 Longhitano parameters & brane localized & ideally delocalized \\
\hline
$e^2 \alpha_1$ & 
$- \frac{2}{3}\, \lambda\, s^2$ &
0\\
\hline
$e^2 \alpha_2$ & 
$  - \frac{1}{12} \left(\frac{7+\kappa}{1+\kappa}\right) \lambda\, s^2 $ &
$  - \frac{1}{12} \left(\frac{7+\kappa}{1+\kappa}\right) \lambda\, s^2 $ \\
\hline
$e^2 \alpha_3$ & 
$  - \frac{1}{12} \left(\frac{1+7\kappa}{1+\kappa}\right) \lambda\, s^2 $ &
$  \frac{1}{12} \left(\frac{7+\kappa}{1+\kappa}\right) \lambda\, s^2 $ \\
\hline
$e^2 \alpha_4$ & 
$ \frac{1}{30}\, \frac{(1+14\kappa+\kappa^2)}{(1+\kappa)^2}\, \lambda\, s^2\ $ &
$ \frac{1}{30}\, \frac{(1+14\kappa+\kappa^2)}{(1+\kappa)^2}\, \lambda\, s^2\ $
\\
\hline
$e^2 \alpha_5$ & 
$-  \frac{1}{30}\, \frac{(1+14\kappa+\kappa^2)}{(1+\kappa)^2}\, \lambda\, s^2\ $&
$ - \frac{1}{30}\, \frac{(1+14\kappa+\kappa^2)}{(1+\kappa)^2}\, \lambda\, s^2\ $
\\
  \hline
 \end{tabular}
\caption{Longhitano's parameters in $SU(2)_A\otimes SU(2)_B$
 flat Higgsless models for brane localized and
 ideally delocalized fermions. }
 \label{table:SU2SU2:f}
\end{table}

These values are consistent with several symmetry considerations.  First, $\alpha_2$ is the coefficient of an operator that is not related to the $SU(2)_W$ properties of the model; as such, this coefficient should be unaffected by the degree of delocalization of the $SU(2)_W$ properties of the fermions.  Indeed, we see that $\alpha_2$ is the same for both the brane-localized and ideal fermions.  Conversely, we expect the values of $\alpha_1$ and $\alpha_3$ to be sensitive to the $SU(2)_W$ properties of the fermions and this is observed in our results, yielding an example of theories in which $\alpha_2 \neq \alpha_3$.   Third, in the limit where $\kappa \to 1$, the models with brane-localized fermions should display an $A \leftrightarrow B$ parity; this is consistent with the fact that $\alpha_2 = \alpha_3$ for $\kappa = 1$.  Finally, since $\Delta\kappa_\gamma\equiv 0$,
we find  $\alpha_2 = - \alpha_3$ for the case of ideal delocalization, in which 
$\alpha_1=0$.

\section{Conclusions}

In this talk, we have seen that Higgsless models are intriguing candidate solutions to the puzzle of Electroweak Symmetry Breaking.  While these models inherently arise from extra-dimensional gauge theories, we find it convenient to employ the technique of deconstruction in order to study the 5-dimensional gauge theories as consistent effective field theories in four dimensions.  We have shown that Higgsless with localized fermions are not phenomenologically viable, because those able to properly unitarize longitudinal electroweak gauge boson scattering have values of $\alpha S$ that are larger than experiment allows.  However, delocalizing the fermions along the 5th dimension can yield viable models; models with ideal delocalization have vanishingly small precision electroweak observables.  The best experimental limits on these models presently come from LEP II bounds on triple-gauge-boson vertices; the mass of the lightest extra W' boson must be at least 500 GeV.  Moreover, the electroweak chiral Lagrangian parameters are calculable in Higgsless models and are of a size that should be accessible to future experiments.


\begin{theacknowledgments}

EHS gratefully acknowledges the kind hospitality of the conference organizers and the support of NSF grant PHY-0354226.

\end{theacknowledgments}


\end{document}